\documentclass[prl,showpacs,twocolumn,superscriptaddress]{revtex4}
\usepackage{amsmath,graphicx,latexsym,bm}

\begin{document}

\title{Enhancement of ferroelectricity at metal/oxide interfaces}

\author{Massimiliano Stengel}
\affiliation{Materials Department, University of California, Santa Barbara,
             CA 93106-5050, USA}
\author{David Vanderbilt}
\affiliation{Department of Physics and Astronomy, Rutgers University, Piscataway,
             New Jersey 08854-8019, USA}
\author{Nicola A. Spaldin}
\affiliation{Materials Department, University of California, Santa Barbara,
             CA 93106-5050, USA}

%\affiliation{Materials Department, University of California, Santa Barbara,
%             CA 93106-5050, USA}

\date{\today}

\begin{abstract}
By performing first-principles calculations on four capacitor
structures based
on BaTiO$_3$ and PbTiO$_3$, we determine the intrinsic interfacial
effects that are responsible for the destabilization of the polar
state in thin-film ferroelectric devices.
We show that, contrary to the established interpretation,
the interfacial capacitance depends crucially on the local
chemical environment of the interface through the
force constants of the metal-oxide bonds, and is
not necessarily related to the bulk screening properties of the
electrode material.
In particular, in the case of interfaces between AO-terminated
perovskites and simple metals, we demonstrate a novel mechanism
of ``interfacial ferroelectricity'' that produces an overall
\emph{enhancement} of the ferroelectric instability of the film,
rather than a suppression as is usually assumed.
The resulting ``negative dead layer'' suggests a route to thin-film
ferroelectric devices that are free of deleterious size effects.

\end{abstract}


\maketitle

%%%%%%%%%%%%%%%%%%%%%%%%%%%%%%%%%%%
\marginparwidth 2.7in
\marginparsep 0.5in
\def\msm#1{\marginpar{\small MS: #1}}
\def\dvm#1{\marginpar{\small DV: #1}}
\def\nsm#1{\marginpar{\small NS: #1}}
\def\scr{\scriptsize}
%uncomment next two lines to have no commentaries
 \def\msm#1{}
 \def\dvm#1{}
 \def\nsm#1{}
%%%%%%%%%%%%%%%%%%%%%%%%%%%%%%%%%%%

%%% INTRO %%%

Capacitors based on ferroelectric perovskites are potentially
attractive for applications in nanoelectronics, such as
non-volatile random-access memories and high-permittivity gate
dielectrics.
Thin-film geometries are sought after for optimal efficiency
and information storage density~\cite{Dawber:2005}.
However, in the thin-film regime, strong size-dependent effects arise.
These can considerably worsen the attractive functionalities of the
ferroelectric material by, e.g., reducing the dielectric
response~\cite{Hwang:2002,Plonka:2005}
and causing rapid polarization relaxation~\cite{Kim_et_al:2005}.
Understanding and addressing these deleterious effects is crucial
for future progress~\cite{Junquera_review:2008}.

Several hypotheses have been formulated to interpret
size effects in thin-film ferroelectric capacitors. Usually
the experimental data can accurately be described in terms of
the ``series capacitor model,'' where the overall capacitance
density $C$ of the device is written as
\begin{equation}
\frac{1}{C} = \frac{1}{C_1} + \frac{4\pi t}{\epsilon_{\rm b}} + \frac{1}{C_2} \,.
\label{eqseries}
\end{equation}
Here $\epsilon_{\rm b}$ is the bulk permittivity of the dielectric
material of thickness $t$ and $C_{1,2}$ are capacitances
associated with low-permittivity layers located at the
film/electrode interfaces.
Such interfacial capacitances produce a depolarizing
field~\cite{Junquera/Ghosez:2003,Pertsev:2007} that strongly
reduces the dielectric response in the paraelectric regime and
can destabilize the single-domain ferroelectric state.
An accurate knowledge of the interfacial capacitance and its
dependence on the combination of ferroelectric and electrode
materials is, therefore, crucial for device design.

Interestingly, such interfacial ``dead layers'' are present
even in high-quality epitaxial systems where the defect
density is very low~\cite{Kim_et_al:2005}.
This suggests possible fundamental issues affecting the
polarization at the metal/ferroelectric boundary.
In particular, within the Thomas-Fermi model, many authors focus on
the imperfect compensation of the polarization charges due to the
finite electronic screening length of metallic electrodes.
To explain the substantially better performance of
SRO electrodes as compared to Pt ones, as has been observed
experimentally~\cite{Plonka:2005,Hwang:2002}, one can then
invoke the lattice contribution to the screening, which
is expected to be significant~\cite{Black/Welser:1999} in oxide
electrodes.

The above interpretation, however, is clearly unsatisfactory at the
microscopic level.
The region where the interface effects occur is as
thin as a few interatomic spacings~\cite{Tagantsev:2006}.
In this region the local chemical and electrostatic environment
departs significantly from that of either parent material, and a
description of the interface in terms of bulk parameters is
unjustified~\cite{Junquera_review:2008}. To describe such effects,
a full quantum-mechanical treatment is required.

First-principles calculations have already been invaluable in
understanding the properties of nanoscale ferroelectrics.
Several studies have focused on the effects of
strain~\cite{Rabe_strain,Ederer/Spaldin_3:2005} and interfacial
electrostatics~\cite{Junquera/Ghosez:2003,Sai/Rappe:2005,Elsasser:2006}
in determining the spontaneous polarization of
symmetric~\cite{Gerra:2006} or asymmetric~\cite{Gerra:2007}
capacitor geometries with various ferroelectric/electrode
combinations.
While most authors agree on the existence of a depolarizing
field that tends to suppress single-domain ferroelectricity,
a clear description of the microscopic mechanisms causing it 
has not yet emerged.
Until it does, it will be difficult to overcome this
deleterious behavior.
The recent development of rigorous first-principles methods to 
define and control the polarization in a metal/ferroelectric
heterostructure~\cite{Stengel/Spaldin:2007,fixedd} now provides the
opportunity to perform such analysis.

In the present work we concentrate on ultrathin ferroelectric
capacitor structures in which the ferroelectric is PbTiO$_3$ (PTO)
or BaTiO$_3$ (BTO) and the metallic electrode can be SrRuO$_3$ (SRO) or Pt.
Aside from being technologically relevant combinations, these are
diverse enough to cover several kinds of behavior while remaining
focused enough to allow an in-depth analysis.
We find that the interfacial dielectric response is more complex
than usually assumed in phenomenological models, where only the
penetration of the electric field in the electrode is considered.
In particular, we develop a rigorous theory,
based on the modern theory of
polarization, in which the interface-specific effects of purely
electronic screening and of interatomic force constants are both
taken fully into account in assessing the overall performance of
the capacitor.
Based on our analysis,
we demonstrate a covalent bonding mechanism that
yields a \emph{ferroelectric} behavior of the interface between
AO-terminated films and simple metals.
\dvm{Italicized ``enhancement''}
In this case one finds an overall {\it enhancement} of the driving force
of the film towards a polar state, rather than a suppression as
is predicted by semiclassical theories.

%{\it Approach.}
%
\dvm{The paper is just long enough that I feel the need for
some internal section labels.  Is there a gentle way of doing
this, consistent with the style of the journal?  Eg, start
a section by starting a paragraph with a short italicized phrase
like this?  If so, I would add a couple more like this later
in the paper...}
Our calculations are performed within the local-density approximation
of density-functional theory and the projector-augmented-wave
method \cite{Bloechl:1994} as implemented in an in-house code.
An isolated capacitor with semi-infinite leads is modeled by
a periodic array of alternating metallic and insulating layers,
where the thicknesses of the metallic slab is treated as a
convergence parameter and the ferroelectric film is $N$ unit
cells thick (the actual values are reported in Table~\ref{tab1}).
We focus on symmetric interface terminations of the
SrO/TiO$_2$ type for SRO electrodes, and of the
Pt$_2$/AO type for Pt electrodes.
In all cases we constrain the in-plane lattice parameter to
the theoretical bulk SrTiO$_3$ value.
We use our recently developed fixed-$D$~\cite{fixedd} method,
combined with the extensions to metal/insulator heterostructures of
Refs.~\onlinecite{Stengel/Spaldin:2007} and \onlinecite{nature_2006}, to
analyze the linear response of the paraelectric structure to
a small polar perturbation.
This regime is relevant for the existence of a (meta)stable
single-domain ferroelectric state, and allows for a microscopic
analysis of the physical ingredients contributing to the screening,
without complications arising from asymmetries or pathological
band alignments~\cite{Junquera_review:2008}.

In order to assess the stability of a given structure against
a polar distortion, we are concerned with $C^{-1}$, its inverse
capacitance per in-plane unit cell, which can be expressed
as~\cite{fixedd}
\begin{equation}
C^{-1} = \Big( \frac{4\pi}{S} \Big)^2 \frac{d^2U}{dD^2} =
   -\frac{4\pi}{S} \frac{dV}{dD}
\label{eqinvcapa}
\end{equation}
where $U$ is the internal energy per in-plane cell, $S$ is the cell
area, and $V$ is the potential drop across the capacitor plates.
We note that $C^{-1}$ is a ``generalized'' inverse capacitance that
is meaningful even when it is negative \cite{fixedd}, in which case
it signals the appearance of an instability to a ferroelectric distortion
in the capacitor arrangement --
the more negative, the stronger the driving force towards a
polar state.

We calculated $C^{-1}$ for the four capacitor heterostructures
considered in this work using a finite-difference approach.
In particular, for each system we performed two complete electronic and
structural relaxations, first at $D=0$ by imposing a mirror
symmetry, and then at $D=0.001$\,a.u.\ (a value that is small enough
to ensure linearity) using our finite-field
technique~\cite{Stengel/Spaldin:2007,fixedd}.
This is possible because the fixed-$D$ method allows one to define and compute
the equilibrium state of an insulating system, within
a given set of symmetry constraints and at a given value of macroscopic
$D$, even in configurations that would be unstable at fixed electric field
(e.g. a ferroelectric near the saddle point of its double-well potential).
Finally, we extracted the induced bias potential and discretized
Eq.~(\ref{eqinvcapa}) to obtain the inverse capacitances, whose values
are reported in Table~\ref{tab1}.
The positive sign of $C^{-1}$ in the PTO/SRO case indicates that
the system is paraelectric for the considered PTO thickness, while the
other structures are in the ferroelectric regime.

\begin{table}
\begin{tabular}{c|c||c|c|c|c}
\hline
 &$N$  & $ C^{-1}S $ (m$^2$/F) &  $C_i^{-1}S$ (m$^2$/F)  & $N_{\rm crit}$ & $\lambda_{\rm eff}$ (\AA)\\
\hline
\hline
BTO/SRO      &    6.5  &  $-$1.553 &   2.280     &   4.85  &  0.202     \\
PTO/SRO      &    6.5  &   \;\;\;0.439 &   1.727     &   7.45   &  0.153    \\
STO/SRO      &    6.5  &   \;\;\;3.876 &   1.647     &    - &  0.146        \\
PTO/Pt       &    8.5  &  $-$1.427 &   1.258     &   5.42  &  0.111     \\
BTO/Pt       &    8.5  &  $-$7.920 &   0.037     &   0.08  &  0.003     \\
\hline
\end{tabular}
\caption{Calculated inverse capacitance densities $C^{-1}S$, interfacial
inverse capacitance densities $C_i^{-1}S$, critical thickness $N_{\rm crit}$
and effective screening length $\lambda_\mathrm{eff}$ for the four capacitor
heterostructures discussed in this work. Data for SRO/STO from
Ref.~\onlinecite{nature_2006} are reported for comparison.}
\label{tab1}
\end{table}

A quantity that has received considerable attention in the recent
past~\cite{Junquera/Ghosez:2003,Sai/Rappe:2005,Gerra:2006,Elsasser:2006}
is the \emph{critical thickness} for ferroelectricity,
defined as the minimum thickness for which a polar %monodomain
state exists.
The usual strategy to determine this quantity computationally is to perform several
calculations for varying thicknesses until a spontaneous polarization
appears.
Instead, we find
it much more convenient to exploit the power of the fixed-$D$
approach and obtain such information directly from a calculation on a
\emph{single} thickness.

\begin{figure}
  \begin{center}
      \includegraphics[width=3.1in]{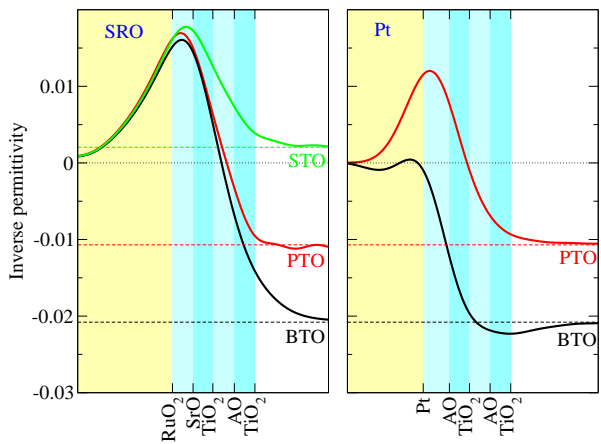}
      \caption{Inverse local permittivity profiles for symmetric capacitor
      structures studied in this work.  (For the SRO electrode, the
      STO curve from Ref.~\onlinecite{nature_2006} is added for
      comparison.)  Inverse permittivities calculated for bulk STO, BTO,
      and PTO are shown as dashed horizontal lines.  Electrodes and
      oxide layers closest to the interface are indicated by shading.
      \label{figperm}}
  \end{center}
\end{figure}

To that end, it is useful to look first at the spatial decomposition of
the inverse capacitances shown in Fig.~\ref{figperm}, where we plot the
local inverse permittivity profiles
$\epsilon^{-1}(x)=d\bar{\cal E}(x)/dD$ (evaluated at $D$=0), where  $\bar{\cal E}$
is the $x$ component of the $y$-$z$--averaged electric field along
the stacking direction $x$.
(These are calculated as in Ref.~\onlinecite{nature_2006}, except
that our use here of the fixed-$D$ method allows treatment of cases
having $C^{-1}<0$.)
The profiles are generally characterized by two regions, an interfacial
part where the imperfect screening manifests itself as a positive peak
in $\epsilon^{-1}(x)$, and the deep interior of the insulating film where
$\epsilon^{-1}(x)$ converges to the bulk value.

This naturally suggests a local decomposition between interface
and bulk effects, where the overall stability of the centrosymmetric
state emerges from the competition between these two usually opposite
contributions.
Rewriting Eq.~(\ref{eqseries}) as $C^{-1}=C_1^{-1} + C_{2}^{-1}
+NC_{\rm b}^{-1}$, we obtain for a symmetric capacitor
\begin{equation}
C^{-1}_i = \frac{C^{-1} - N C^{-1}_{\rm b}}{2} \, .
\label{intcap}
\end{equation}
Here $N$ is the number of bulk cells and
$C_{\rm b}^{-1}=4\pi c_{\rm b}/ (\epsilon_{\rm b} S)$ is the bulk
inverse capacitance per unit cell ($c_{\rm b}$ is the cell parameter
along the field direction).
We take Eq.~(\ref{intcap}), based on a symmetric capacitor,
to be our \emph{definition} of the interfacial inverse capacitance.
(Note that the definition
depends on the precise convention for specifying $N$; we adopt the
convention that $N$ is the nominal thickness as illustrated by the
examples in Table \ref{tab1}.)
Then, given $C^{-1}_i$ and $C^{-1}_{\rm b}$, we again use Eq.~(\ref{eqseries})
to \emph{predict} the critical thickness
\begin{equation}
N_{\rm crit} = -\frac{C_1^{-1} + C_{2}^{-1}}{C_{\rm b}^{-1}}
\label{eqncrit}
\end{equation}
for ferroelectricity in a (possibly asymmetric) capacitor,
this being the value of $N$ that yields
an overall vanishing inverse capacitance.
For an interface to a ferroelectric material ($C_{\rm b}^{-1}<0$), it is
also natural to define the effective \emph{dead-layer
thickness} $N_{{\rm dead},i}=-C_i^{-1}/C_{\rm b}^{-1}$, in terms of which
\begin{equation}
N_{\rm crit} = N_{{\rm dead},1}+N_{{\rm dead},2}
\label{eqncritdead}
\end{equation}
or, for a symmetric capacitor, $N_{\rm crit} = 2N_{\rm dead}$.

The calculated values for $C_i^{-1}$ and $N_{\rm crit}$ are reported
in Table~\ref{tab1}, together with the values already calculated in
Ref.~\cite{nature_2006} for paraelectric SrTiO$_3$ (STO) between
SRO electrodes.
Since many authors discuss interfacial effects in terms of
effective screening lengths $\lambda_\mathrm{eff} = C^{-1}_i S / 4\pi$
rather than capacitances, we also report these values in
the table.

It is rather surprising that the calculated $N_{\rm crit}$ of BTO/SRO
is smaller than that of PTO/SRO.
Indeed, PTO has a much larger spontaneous polarization
$P_s$ in the tetragonal ground state than BTO, and one would be
tempted to think that the former is a ``stronger'' ferroelectric
and therefore should have a smaller critical thickness.
The calculated trend in $N_{\rm crit}$ is even more surprising on
observing that $C_i^{-1}$ of PTO/SRO is smaller
than that of BTO/SRO, which would favor the opposite trend.
It is clear from Eq.~(\ref{eqncrit}), however, that $|C_{\rm b}^{-1}|$,
rather than $P_s$, is the parameter that defines the strength
of the bulk ferroelectric instability; at the fixed STO in-plane
lattice constant, our calculated $|C_{\rm b}^{-1}|$ is about twice as large
in BTO than in PTO, which agrees nicely with the trend in $N_{\rm crit}$.
This highlights the fact that $|C_{\rm b}^{-1}|$, rather than $P_s$, is
the important parameter to consider when discussing the competition
of bulk and interfacial electrostatics in ferroelectric capacitors.

\begin{figure}
  \begin{center}
      \includegraphics[width=3.2in]{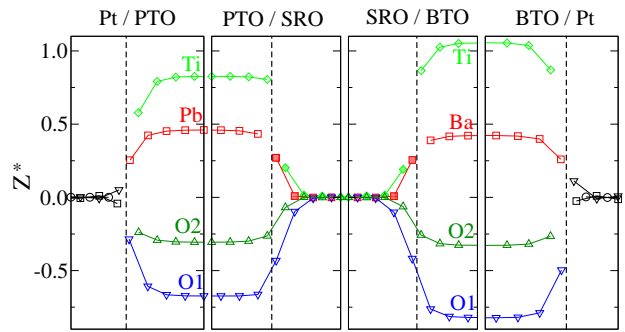} \\
      \vspace{5mm}
      \caption{Plots of Callen dynamical charges, arranged head-to-head
      and tail-to-tail to emphasize rapid convergence to common bulk values.
      Colored symbols denote A-site cations (red squares),
      B-site cations (light green diamonds), AO oxygens (O1, blue down triangles),
      and BO$_2$ oxygens (O2, dark green up triangles). The Pt atoms 
      (black symbols) are labeled according to their projection onto the
      interface plane. Of the four independent Pt sites, one is shared with
      the A cations (squares), another is shared with O1 (down triangles) and 
      the remaining two, equivalent by symmetry, are shared with O2 
      (circles).
       \label{figzstar}}
  \end{center}
\end{figure}

\msm{\scr Caption of Fig.~\ref{figzstar}:
I've been struggling to find a reasonable wording. I tried a rewrite, see what
you think.}
\dvm{Fine.}

In the Pt-based systems considered here, $\lambda_\mathrm{eff}$ is systematically
smaller than in the SRO-based systems, in stark contrast with
the predictions of the semiclassical models of
Refs.~\onlinecite{Black/Welser:1999,Kim_boundcharges}.
(First-principles calculations have already demonstrated, in the case of STO
paraelectric capacitors, that Pt electrodes can be intrinsically superior
to SRO electrodes~\cite{nature_2006}.)
Here, while in PTO/Pt $\lambda_\mathrm{eff}$ is still significant, 
in the case of BTO/Pt it practically vanishes,
or equivalently, the dead-layer thickness of $N_{\rm dead}=0.04$ layers
is almost zero!
This suggests that novel effects must take place at this interface that
go beyond the usual Thomas-Fermi arguments, which always predict
the same, positive value independent of the ferroelectric material.

To investigate the origin of such effects, we perform a microscopic
analysis of the electrostatic response to a $D$ field in the two
Pt-based capacitors.
The static inverse capacitance can be decomposed as
\begin{equation}
C^{-1} = (C^\infty)^{-1} - \Delta
\label{eqcapa}
\end{equation}
where $(C^\infty)^{-1}$ is the purely
electronic, frozen-ion value and $-\Delta$ is the lattice contribution.
The latter can be expressed in terms of the
Callen (or longitudinal) dynamical charge tensor $Z^L$ \cite{Ghosez_etc},
and the longitudinal force constant matrix $\mathbf{K}^L$, as
\begin{equation}
\Delta = \Big( \frac{4\pi}{S} \Big)^2 \sum_{ij} Z^L_{i,x}
(K^L)^{-1}_{ij} Z^L_{j,x}
\label{eqdeltak}
\end{equation}
where $x$ is again the stacking direction.
Note that the sums in Eq.~(\ref{eqdeltak}) run over \emph{all} atoms in
principle, including those located deep in the metallic electrodes,
but in practice these do not contribute because their
dynamical charges are zero (see Fig.~\ref{figzstar}).
The convenience of using Eq.~(\ref{eqdeltak}) instead of the usual
decomposition involving transverse quantities, e.g. Eq.~(8) of
Ref.~\onlinecite{Antons:2005}, is that, in a layered heterostructure,
$\mathbf{K}^L$ is \emph{short-ranged} in real
space~\cite{Giustino_ir,Xifan_sl}.
This means that local bonding effects can be unambiguously
separated from the long-range electrostatic interactions, thus allowing
for a detailed analysis of the microscopic mechanisms contributing
to the polarization.

It is clear from Eq.~(\ref{eqdeltak}) that two basic ingredients contribute
to the ionic polarizability, the force constants contained in $\mathbf{K}^L$,
and the dynamical charge associated with a given degree of freedom, $Z^L$.
Since the \emph{inverse} of $\mathbf{K}^L$ enters the sum, it is obvious
that \emph{less stable} bonds will yield an enhanced response, provided that
the participating atoms carry a significant $Z^L$.
It is reasonable then to expect that the relative weakness of the bonds
formed by the ferroelectric films with Pt, compared to those formed with
the isostructural conducting oxide SRO, might explain the enhanced
interfacial dielectric properties of the former electrode material,
and might also help shed some light on the qualitatively different
behavior of PTO/Pt and BTO/Pt.

\begin{figure}
\begin{center}
\includegraphics[width=3.0in]{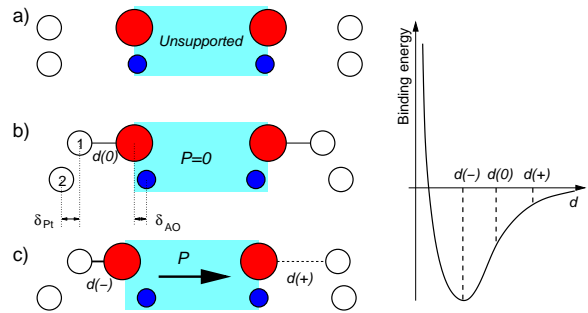}
\end{center}
\caption{Schematic representation of the salient features of
 the Pt$_2$-AO interface. Only interfacial atoms are shown; empty, red, and blue
 circles denote Pt, O, and A atoms respectively.
 (a) Unsupported film and metal surfaces. A small surface rumpling is
 present only in the AO surface layer while the metal is flat.
 (b) Centrosymmetric interface geometry, with significant
 rumplings $\delta$, especially in the Pt$_2$ layer.
 (c) Proposed mechanism for interfacial ferroelectricity,
 driven by the unstable Pt-O bond.
 Right: example of a two-body binding energy curve having
  negative curvature at $d(0)$.
 \label{figsketch}}
\end{figure}

\begin{table}
\begin{center}
\begin{tabular}{c||c|c|c}
\hline
System & $d$ (\AA)  & $\delta_\mathrm{Pt}$ (\AA) & $\delta_\mathrm{AO}$ (\AA) \\
\hline
BTO/Pt & 2.26 &      0.46            &     0.12             \\
PTO/Pt & 2.11 &      0.34            &     0.19             \\
\hline
\end{tabular}
\end{center}
\caption{Computed structural parameters for Pt$_2$-AO interfaces.
See Fig.~\ref{figsketch} for the definition of the symbols.
\label{tab2}}
\end{table}

\dvm{Table II: I think you need to put the units in the top row after
each symbol, like ``$d$ (a.u.)'' (if it is a.u.).  Regarding formatting,
there is some revtex4 command to stretch a table to column width; I don't
remember it at the moment, but you might want to use it.  (Also,
I'm very conditioned by Phys Rev journals; they would tell you
to take most or all of the vertical lines out of the table (the same
applies to Table I).)}

To investigate this hypothesis, we start by looking at the relaxed geometry
of the interfacial Pt$_2$ and AO layers when the film is in the
centrosymmetric configuration, schematically shown in Fig.~\ref{figsketch}.
The significant rumpling $\delta$ which is present in both Pt$_2$ and AO
(values are reported in Tab.~\ref{tab2}) brings Pt(1) and the interface
O atoms in much closer contact than the neighboring Pt(2) and A cation,
indicating that only the former two atoms form a true chemical bond.
Moreover, the strong buckling in the Pt$_2$ layer is indicative of a
frustrated bonding environment, where the repulsive A-Pt interaction is
in competition with the attractive O-Pt bond.
As a result, the O-Pt distance $d$ in the relaxed interface structures is in
both cases significantly larger than the typical Pt-O equilibrium distance
of about 2.0\,\AA, which has been determined experimentally and theoretically in
a number of bulk oxide phases~\cite{Ciacchi_PtO}.
Comparing the interface systems, $d$ is even larger in BTO/Pt
than in PTO/Pt, indicating that the Pt-O bond is weaker in the former
system.

To quantify this difference in strength, we examine the local force
constants characterizing the Pt-O interfacial bond. At the
PTO/Pt interface we have $K^L_{\mathrm{Pt-O}} = -0.039$\,a.u., which is
comparable to the typical interatomic force constants in oxide crystals
and is consistent with a \emph{stable} chemical bond.
Strikingly, at the BTO/Pt interface we have
$K^L_{\mathrm{Pt-O}} = 0.001$\,a.u., a very small and
positive value, which indicates that the Pt-O bond here is slightly
\emph{unstable} (i.e. there is no restoring force associated
to it).
This indicates that the structural frustration present at the interface
is strong enough in BTO/Pt to pull the Pt-O bond into an unstable regime
[see Fig. 3 (right)].
This qualitative difference between PTO/Pt and BTO/Pt is entirely
responsible for their inequivalent polar response, as we shall see
in the following.

\begin{figure}
  \begin{center}
      \includegraphics[width=3.1in]{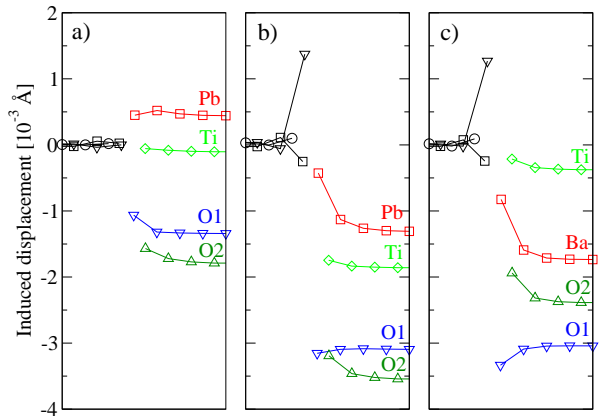} \\
      \vspace{0.2in}
      \caption{Ionic displacements induced by a small field of
      $D$ = 10$^{-3}$\,a.u.\ (0.455\,$\mu$C/cm$^2$).
      (a) PTO/Pt; (b) PTO/Pt modified by artificial addition of force
      constant $k=-0.05$\,a.u.; (c)
      BTO/Pt. The color/symbol scheme is the same as in Fig.~\ref{figzstar}.
      The horizontal axis is the $x$ coordinate, normal to the interface plane, and
      refers to the relaxed atomic positions in the centrosymmetric geometry.
      \label{figdx}}
  \end{center}
\end{figure}

In order to link the stability of the Pt(1)-O bond to the dielectric
response of the capacitor, we perform a computational experiment in
which we artificially modify the force constant matrices $\mathbf{K}^L$
of PTO/Pt and BTO/Pt by adding a negative harmonic term
of the form $E_h= (k/2)  [x_\mathrm{O}-x_\mathrm{Pt}-d(0)]^2$ between Pt(1)
and O, which weakens the corresponding bond without changing the
centrosymmetric equilibrium geometry.
Our goal is to demonstrate that, by choosing an appropriate value of $k$,
one can destabilize the Pt(1)-O bond at the PTO/Pt interface and therefore
reproduce the behavior (dead-layer thickness and atomic displacement pattern)
of the BTO/Pt system.
To that end, we choose $k=-0.05$\,a.u.\ (a value slightly larger
than $K^L_{\mathrm{Pt-O}}$) and we recalculate the effective
dead-layer thickness for PTO/Pt by inserting this modified
$\mathbf{K}^L$ matrix into Eq.~(\ref{eqdeltak}).
Remarkably, due to the additional harmonic term $E_h$, $N_\mathrm{dead}=2.71$
jumps to a very small value $N_\mathrm{dead}=0.08$, practically
identical to that of the unmodified BTO/Pt case.

To further prove that the ``modified'' PTO/Pt behaves in all important respects
like BTO/Pt, we plot in Fig.~\ref{figdx} the induced ionic
displacements in three different cases: PTO/Pt ($k=0$),
PTO/Pt ($k=-0.05$), and BTO/Pt ($k=0$).
Note that the ionic responses to a small $D$ field are very different in
the first and last cases.
In this context, it is remarkable that the middle ``modified PTO/Pt'' case
is strikingly similar to BTO/Pt, even in many fine details of the induced 
relaxations at the interface layers.
Deeper into the oxide, the displacements reflect the different
ferroelectric mode patterns (e.g., there is a larger displacement for
Pb than for Ba because of the stereochemical activity of the Pb lone
pairs), but the behavior of the first oxide layers is nearly
identical.
This unambiguously demonstrates that the dissimilar behavior in the
BTO/Pt and PTO/Pt cases is almost entirely due to the change in
stability of the Pt-O bond.

Now, to have a mechanism for ferroelectricity, a structural instability
needs to be coupled to the polarization.
Comparing the original and the ``modified'' PTO/Pt ionic displacement
patterns in Fig.~\ref{figdx}(a) and (b), it is apparent that the
stretching of the Pt-O bond is associated with i) a shift of the
whole ferroelectric slab in the opposite direction to $D$,
and ii) with a combined rumpling of the AO and Pt$_2$ layers.
Since the atoms in both Pt$_2$ and AO carry significant values of
$Z^L$ (see Fig.~\ref{figzstar}), both effects have a strong impact on
the polarization.
We therefore identify this mechanism as a novel form of ``interfacial
ferroelectricity.'' Here we have a flagrant breakdown of
phenomenological models based on bulk properties of the parent
materials.

\begin{figure}
  \begin{center}
      \includegraphics[width=3.1in]{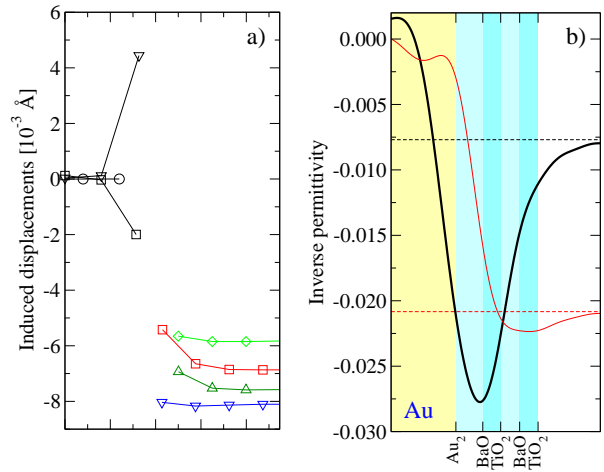}
      \caption{Computed local properties at the BTO/Au interface. (a) Ionic displacements
      induced by a small field of $D$ = 0.001\,a.u.; (b)
      Inverse permittivity profile (thick solid black curve) compared with
      corresponding BTO/PT results (thin solid red curve) 
      as in Fig.~\ref{figperm}.
      The thin dashed lines in (b) indicate the values of the bulk inverse 
      permittivity of BTO at $D=0$ for two different values of the in-plane 
      strain, 0\% (black) and $-$2.2\% (black), corresponding to the cubic
      lattice constants of BTO and STO, respectively.
      \label{figau}}
  \end{center}
\end{figure}

These arguments lead to a quite general prediction, namely that AO-terminated
perovskite ferroelectrics can show a strong interfacial
enhancement of the ferroelectric properties when weakly bonded to
a simple metal.
To test this prediction, we performed a supplementary calculation
for a BTO-based capacitor by replacing the Pt electrodes with a
more inert material, Au.
In this case the equilibrium lattice constant of bulk BTO was used for the
epitaxial strain constraint, to better match the experimental
setup of Ref.~\cite{Saad/Scott:2004}, but otherwise all
computational parameters were kept the same.
As before, we first relaxed the centrosymmetric geometry and
then calculated the ionic and electronic response to a small field.
The calculated induced ionic displacements for BTO/Au, presented
in Fig.~\ref{figau}(a), show the same qualitative features i) and ii) discussed
above, but their magnitude is about four times larger than in
BTO/Pt, reflecting the weaker Au-oxide interaction.
Consistent with the enhanced displacements, we obtain a \emph{negative
dead-layer thickness} of $N_{\rm dead}=-5.5$ layers, which manifests itself
as an unusually large dip in the local permittivity profile, shown in
Fig.~\ref{figau}(b).
(Note the rather dissimilar bulk ``ferroelectric
strength'' of BTO at the different in-plane strain states.) 
In other words, the actual capacitor with 8.5 layers of BTO and real
Au electrodes has the same degree of ferroelectric instability as
$\sim$19.5 layers of BTO between ``ideal'' electrodes!
It is quite remarkable that such a
huge effect can be caused by only two atomic monolayers
at the interface (Au$_2$ and BaO).

In conclusion, we have used a rigorous finite-field approach to
study the influence of the electrode-film interface on the
ferroelectric instability of nanoscale BaTiO$_3$ and PbTiO$_3$ capacitors.
At the AO-terminated interfaces with simple metals, we demonstrate
a strong correlation of the overall ferroelectric response
of the capacitor to the strength of the interfacial metal-oxygen bond.
In cases where this bond is especially weak, such as the BaO-Au interface,
we find an unusually large enhancement of the ferroelectric
instability.
This result is in striking contrast with the conclusions of
the Thomas-Fermi screening model, demonstrating that a microscopic
analysis is generally necessary to describe interfacial effects in
ferroelectric capacitors.
This mechanism also suggests a possible route to the fabrication of
nanoscale devices free from interfacial size effects.

\section*{Methods}

Our calculations are performed within the local-density approximation
of density-functional theory and the projector-augmented-wave
method \cite{Bloechl:1994}, with a planewave basis cutoff energy
of 40 Ry.
We constrain the in-plane lattice constant to the calculated
theoretical equilibrium value for cubic SrTiO$_3$ ($a$ = 7.276
a.u.) unless otherwise specified.
We set the thickness of the insulating film to $N$ unit
cells (the actual values are reported in Table~\ref{tab1}).
The thickness of the metal electrode slab was
7.5 unit cells of SRO or 11 atomic layers of Pt, which were
sufficient to converge the interface properties of interest.
We use a $6 \times 6$ Monkhorst-Pack sampling of the
surface Brillouin zone, which reduces to 6 special points
within the tetragonal symmetry. A Gaussian smearing
of 0.15 eV was used to accelerate convergence of the Brillouin-zone
integrations.
We carefully checked that the band alignment at the interface
and the choice of the smearing scheme is such that any spurious
population of the conduction band of the insulating film is
avoided.
All the structural degrees of freedom were fully relaxed
subject to the translational and point symmetries
(a mirror symmetry plane is first imposed to relax the
geometry in the paraelectric state, and later relaxed when
calculating the linear response to a displacement field).

The insulating character of the system along the stacking direction
(the insulator is thick enough so that direct tunneling is
suppressed) allows for a rigorous definition of the macroscopic
polarization and its coupling to an external bias
voltage~\cite{Stengel/Spaldin:2007,nature_2006}.
Because at the quantum-mechanical level
all wavefunctions are mutually coupled and there is no clear way to
separate free conduction charge from the bound polarization charge
we adopt the convention of treating \emph{all} charges as bound 
charges, so that $D$ is a constant throughout the
heterostructure~\cite{explan-longit}
just as for a purely insulating system~\cite{Giustino/Pasquarello:2005}.
What would customarily be called the macroscopic
free charge $\sigma$ on the top of the electrode is
here reinterpreted as a bound charge
associated with a polarization $P=\sigma$ in the metallic region.
(Indeed $D=4\pi\sigma$,
since the electric field must vanish in the metallic region.)

All quantities requiring a derivative with respect to $D$ were performed
by finite differences between a calculation with mirror symmetry imposed
($D=0$) and a fixed-$D$ calculation with $D=0.001$\,a.u.
The longitudinal force constant matrices and the dynamical
charges were computed by finite differences by taking displacements
of $0.4$ m\AA{} along the tetragonal axis.

\section*{Acknowledgements}

This work was supported by the Department of
Energy SciDac program on Quantum Simulations of Materials and
Nanostructures, grant number DE-FC02-06ER25794 (M.S. and N.S.),
and by ONR grant N00014-05-1-0054 (D.V.).

\bibliographystyle{naturemag}
\bibliography{Max}

\begin{thebibliography}{10}
\bibitem{explan-longit}
\bibinfo{note}{Strictly speaking, $D$ and $P$ here are the longitudinal
  components of the respective fields}.
\end{thebibliography}

\end{document}